
\input phyzzx

\def\TableOfContentEntry{}
\def\Roman#1{\uppercase\expandafter{\romannumeral #1}.}
\sectionstyle={\Roman}
\def\section#1{\par \ifnum\the\lastpenalty=30000\else
   \penalty-200\vskip\sectionskip \spacecheck\sectionminspace\fi
   \global\advance\sectionnumber by 1
   \xdef\sectionlabel{\the\sectionstyle\the\sectionnumber}
   \wlog{\string\section\space \sectionlabel}
   \TableOfContentEntry s\sectionlabel{#1}
   \noindent {\caps\uppercase\expandafter{\romannumeral\the\sectionnumber
.}\quad #1}\par
   \nobreak\vskip\headskip \penalty 30000 }

\def\subsection#1{\par
   \ifnum\the\lastpenalty=30000\else \penalty-100\smallskip \fi
   \noindent\enspace{#1}\enspace \vadjust{\penalty5000}}

\font\titlefont=cmcsc10 at 14pt
\font\namefont=cmr12
\font\placefont=cmsl12
\font\abstractfont=cmbx12

\font\sxrm=cmr6          \font\sxmi=cmmi6             
  \font\sxsy=cmsy6           
  
\font\nnrm=cmr9          \font\nnmi=cmmi9
  \font\nnsy=cmsy9           \font\nnex=cmex10 at 9pt
  \font\nnit=cmti9           \font\nnsl=cmsl9
  \font\nnbf=cmbx9
\def\nnpt{\def\rm{\fam0\nnrm}%
  \textfont0=\nnrm \scriptfont0=\sxrm \scriptscriptfont0=\fiverm
  \textfont1=\nnmi \scriptfont1=\sxmi \scriptscriptfont1=\fivei
  \textfont2=\nnsy \scriptfont2=\sxsy \scriptscriptfont2=\fivesy
  \textfont3=\nnex \scriptfont3=\nnex \scriptscriptfont3=\nnex
  \textfont\itfam=\nnit \def\it{\fam\itfam\nnit}%
  \textfont\slfam=\nnsl \def\sl{\fam\slfam\nnsl}%
  \textfont\bffam=\nnbf \def\bf{\fam\bffam\nnbf}%
  \normalbaselineskip=11pt
  \setbox\strutbox=\hbox{\vrule height8pt depth3pt width0pt}%
  \let\big=\nnbig \let\Big=\nnBig \let\bigg=\nnbigg \let\Bigg=\nnBigg
  \normalbaselines\rm}
\skip\footins=0.2in
\dimen\footins=4in
\catcode`\@=11
\def\vfootnote#1{\insert\footins\bgroup\nnpt
  \interlinepenalty=\interfootnotelinepenalty
  \splittopskip=\ht\strutbox
  \splitmaxdepth=\dp\strutbox \floatingpenalty=20000
  \leftskip=0pt \rightskip=0pt \spaceskip=0pt \xspaceskip=0pt
  \parfillskip=0pt plus 1fil
  \setbox1=\hbox{*}\parindent=\wd1\let\enspace=\null
  \hangafter1\hangindent\parindent\textindent{#1}\footstrut
  \futurelet\next\fo@t}
\catcode`\@=12
\def\footnoterule{\kern-3pt \hrule width2truein \kern 3.6pt}

\footline={\iftitlepage\hfil\global\titlepagefalse
\else\hss\tenrm\folio\hss\fi}

\newif\iftitlepage

\def\center{\parindent=0pt\leftskip=1in plus 1fill\rightskip=1in plus 1fill}

\def\preprint#1//#2//#3//{\baselineskip=14pt{\rightline{#1}\par}\vskip
.1in{\rightline{#2}\par} {\rightline{#3}}\medskip}

\def\preprintb#1//#2//{\baselineskip=14pt{\leftline{#1}\par}{\leftline{#2}}}

\def\title#1\par{{\center\baselineskip=16pt
  \titlefont{#1}\par}}
\long\def\author #1// #2// #3// #4//{{\baselineskip=14pt\parskip=0pt
   \center\namefont{#1}\par\placefont#2\par#3\par#4\par}\par}

\long\def\abstract#1//{%
  \centerline{\abstractfont Abstract}\par
  {\baselineskip=14.5pt\advance\leftskip by 0pc\advance\rightskip by
0pc\parindent=10pt
  \def\enspace{\kern.3em}
  \noindent #1\par}}

\newcount\refno\refno=0
\long\def\rfrnc#1//{\advance\refno by 1
  $ $\llap{\hbox to\leftskip{\hfil\the\refno.\enspace}}#1\par\medskip}
\def\PL{{\it Phys.~Lett.}}
\def\NP{{\it Nucl.~Phys.}}
\def\PR{{\it Phys.~Rev.}}
\def\PRL{{\it Phys.~Rev.~Lett.}}

\def\bib#1{\hbox{${}^{#1}$}}

\def\t1{{\tilde 1}}

\def\a3{\alpha_3(m_{Z^0})}

\def\GeV{\,{\rm GeV}}
\def\TeV{\,{\rm TeV}}

\newdimen\jbarht\jbarht=.2pt
\newdimen\vgap\vgap=1pt
\newcount\shiftfactor\shiftfactor=12

\catcode`\@=11
\def\jbarout{\setbox1=\vbox{\offinterlineskip
  \dimen@=\ht0 \multiply\dimen@\shiftfactor \divide\dimen@ 100
    \hsize\wd0 \advance\hsize\dimen@
  \hbox to\hsize{\hfil
    {\multiply\dimen@-2 \advance\dimen@\wd0
    \vrule height\jbarht width\dimen@ depth0pt}%
    \hskip\dimen@}%
  \vskip\vgap\box0\par}\box1}
\def\jbar#1{\mathchoice
  {\setbox0=\hbox{$\displaystyle{#1}$}\jbarout}%
  {\setbox0=\hbox{$\textstyle{#1}$}\jbarout}%
  {\setbox0=\hbox{$\scriptstyle{#1}$}\jbarout}%
  {\setbox0=\hbox{$\scriptscriptstyle{#1}$}\jbarout}}
\catcode`\@=12

\def\simlt{\mathop{\lower.4ex\hbox{$\buildrel<\over\sim$}}}
\def\simgt{\mathop{\lower.4ex\hbox{$\buildrel>\over\sim$}}}


\newcount\refno\refno=0
\long\def\rfrnc#1//{\advance\refno by 1
  $ $\llap{\hbox to\leftskip{\hfil\the\refno.\enspace}}#1\par\smallskip}
\def\PL{{\it Phys.~Lett.}}
\def\NP{{\it Nucl.~Phys.}}
\def\PR{{\it Phys.~Rev.}}
\def\PRL{{\it Phys.~Rev.~Lett.}}

\def\bib#1{\hbox{${}^{#1}$}}

\def\t1{{\tilde 1}}

\def\a3{\alpha_3(m_{Z^0})}

\def\GeV{\,{\rm GeV}}
\def\TeV{\,{\rm TeV}}



%

\def\ds{$\delta\tilde m^2_{\jbar d_As_B}$}
\def\db{$\delta\tilde m^2_{\jbar d_Ab_B}$}
\def\leaderfill{\leaders\hbox to 1em{\hss\rm .\hss}\hfill}

\advance\vsize by .25 truein 
\hoffset.25in
\hsize6in
\titlepagetrue

\preprint{}//{MIU-THP-93/65}//{September, 1993}//
\vskip .5truein
\title Testing Unified Supersymmetric Models

\bigskip
\author  S. Kelley//
\vskip.1in Department of Physics//
Maharishi International University//
Fairfield, IA  52557-1069, USA//

\bigskip
\abstract

I define the Standard Supersymmetric Model (SSM) as the minimal supersymmetric
extension of
the Standard Model with gauge coupling unification and universal  soft
supersymmetry
breaking at the unification scale.  This well-defined model has a
five-dimensional space of
unknown parameters ($m_t, \tan\beta, m _{1/2}, m_0, A$).  I outline the
top-down and
bottom-up methods of solving this model.  Thresholds may be treated either by
dropping heavy
particles from the RGE's or by considering loop corrections to the vacuum
energy.
Substantial regions of the parameter space are consistent with all experimental
constraints.  I consider the relation of the SSM to more realistic models such
as
supersymmetric $SU(5)$, Flipped
$SU(5)\times U(1)$, the String-Inspired Standard Model, and string-derived
models. I briefly
discuss sparticle spectroscopy and flavor changing neutral currents (FCNC's) as
sample
methods for determining the unknown parameters of the SSM, and discriminating
between the
SSM and more realistic models. //
\vskip .6in
\centerline{Invited talk at {\it Recent Advances in the Superworld}}
\centerline{Houston Advanced Research Center, April 13--16, 1993}

\vskip .5in
\preprintb{MIU-THP-93/65}//{September, 1993}//
\endpage

\advance\vsize by -1truein 
\normalspace
\pagenumber 1
\section{Introduction} Unified supersymmetric gauge theories are very likely
the framework
to describe physics below the Planck scale.  In addition to the theoretical
motivation that
supersymmetry solves the gauge hierarchy problem, the excellent agreement of
LEP
data with the predictions of gauge coupling constant unification in
supersymmetric models\bib{1} provides a strong basis for belief in
supersymmetry.  Recent precision LEP measurements have sparked an explosion of
work on unified
supersymmetric models.  In sorting through the many different models
available, the elucidation of a Standard Supersymmetric Model (SSM), comparable
to the Standard Model in definiteness and number of parameters, proves useful.
The
SSM is defined as the minimal extension of the Standard Model with gauge
coupling
unification and universal soft supersymmetry breaking at the unification scale.
The SSM has an unknown parameter space of five dimensions, ($m_t, \tan\beta, m
_{1/2},
m_0, A$).

In this talk, I present the basic properties of the SSM, the techniques used to
extract predictions over the complete five-dimensional parameter space, the
relation of the SSM to more realistic models, and discuss sparticle
spectroscopy
and flavor changing neutral currents as sample tests of unified supersymmetric
models.  Section 2 defines the SSM, and Section 3 outlines the top-down and
bottom-up methods of solving the SSM. Section 4 considers how thresholds may be
treated by dropping the heavy particles from the RGE's or considering loop
corrections to the vacuum energy. Section 5 considers the relation of the SSM
to more
realistic models.  Two sample experimental tests of unified supersymmetric
models,
sparticle spectroscopy and flavor changing neutral currents (FCNC's), are
discussed in Sections 6 and 7.  Conclusions are presented in Section 8.

Unification and supersymmetry go hand in hand.  Supersymmetry ensures the
cancellation of quadratic divergences necessary to reliably connect
unification-scale physics with accessible low-energy predictions.  Unification
ensures that the model has only a few parameters, introducing
correlations between the observables of the low-energy model which can be
experimentally tested.  Because of quantum loop-effects, every prediction of
a unified supersymmetric theory is sensitive to all the fundamental parameters
of
the theory.  In principle, any five exact experimental measurements determine
the
parameters of the SSM, and each additional measurement tests the theory.  In
practice, uncertainty in experimental measurements gives an allowed region in
parameter space which narrows, and possibly disappears, with increasing
experimental
precision.

A strict definition of a valid scientific theory requires that the theory be
falsifiable.  The SSM could be falsified in two ways, observing processes like
proton decay or lepton flavor violation which are forbidden in the SSM, or
showing that every point in the parameter space is inconsistent with some
experimental measurement.  Thus it is important to develop general methods for
accurately extracting the predictions of the SSM over its entire parameter
space.

Theoretical considerations clearly indicate that the SSM is only
an approximation to the complete theory of nature.  Moreover, quantum
gravitational effects probably introduce an uncertainty into every prediction
of
the SSM.\bib{2}  However, the ability to test the SSM without
non-renormalizable corrections
from quantum gravity is the foundation for investigating these corrections.
Because of this,
precision calculations and experimental tests of the SSM are crucial.  The
definiteness and
simplicity of the SSM make it the ideal example for developing tools to test
unified
supersymmetric models.  These tools are generally applicable to more realistic
models, but
easier to develop in the context of a simple, well-defined model.

String theory provides a framework to quantify the effects of quantum gravity,
and
realistic string models reduce to unified supersymmetric models below the
Planck scale.  Therefore, the techniques to test the SSM are a necessary
subset of those needed to test string theories.  Also, because of the huge
diversity of possible string models, the field-theoretic parameterization
afforded by string-inspired models is a necessary complement to the ``needle
in the haystack" approach of specific string-derived models.

The SSM should be taken seriously as a well-defined, testable model, which
incorporates most of the features the complete theory of nature is likely
to contain.  The methods of extracting the predictions of the SSM and the
process of experimentally testing it are thus extremely important.
However, the main goal of testing the SSM is to discriminate the fine details
of what lies beyond.

\section{The SSM}

The SSM is defined as the minimal supersymmetric extension of the Standard
Model
with gauge coupling unification and universal soft supersymmetry breaking at
the
unification scale.  The SSM superpotential
$$W=D^C_I\lambda^{IJ}_1Q_jH+U^C_I\lambda^{IJ}_2 Q_J\jbar
H+E^C_I\lambda^{IJ}_3L_JH+\mu H\jbar H\eqno{(2.1)}$$
involves a dimensionful Higgs mixing term, $\mu$. By adding a gauge singlet to
the model
and replacing the last term in the superpotential Eq.~(2.1) with
$$W_{singlet}=\lambda_7\phi H\jbar H+\lambda_8\phi^3\eqno{(2.2)}$$
the Higgs mixing can be
dynamically generated without a dimensionful coupling.\bib{3}  Although the
final,
unconstrained, five-dimensional parameter space of the model  with a
dimensionful Higgs mixing and the model with an extra singlet are identical,
the structure, solution, and constraints differ between the two models.
The analysis of the model defined by Eq.~(2.2) is computationally more complex.
 In what
follows, only the superpotential Eq. (2.1) will be considered.

The Yukawa couplings $\lambda_1,\lambda_2,\lambda_3$ in Eq. (2.1) are general
$3\times 3$
matrices in generation space.  A superfield basis in flavor space which keeps
the gauge
couplings diagonal may be chosen so that
$$\lambda_1=\lambda_d\quad\lambda_2=\lambda_uK\quad
\lambda_3=\lambda_e\eqno{(2.3)}$$
where $\lambda_d$, $\lambda_u$, $\lambda_e$ are
diagonal matrices with real entries and $K$ is the KM matrix.  This
generational
structure of the Yukawa couplings underlies attempts to understand the observed
fermion masses and mixings,\bib{4} and calculation of supersymmetric
contributions to
FCNC's.  However, for most other applications of the SSM, taking a unit KM
matrix
and neglecting all but the top, bottom, and tau Yukawas is an excellent
approximation
which will be used throughout this paper, except for the section on FCNC's.

The boundary conditions for the model at the unification scale are particularly
simple,
involving only ten parameters:  the unification scale $M_X$, the unified gauge
coupling, $\alpha_X$, the third generation Yukawas $\lambda_b$, $\lambda_t$,
$\lambda_{\tau}$,  the universal soft supersymmetry-breaking parameters,
$m_{1/2}$, $m_0$,
$A$, the supersymmetric Higgs mixing parameter $\mu$, and the soft
supersymmetry-breaking
Higgs mixing parameter $B$.  From these parameters renormalized at $M_X$, the
coefficients
in the SSM Lagrangian at any lower renormalization point may be determined
through the
RGE's.\bib{5}  Compared to a theory with general, unrelated soft
supersymmetry-breaking
gaugino masses, scalar masses, and trilinear couplings, this is an immense
reduction in the
number of parameters.\bib{6} The known experimental values of $m_b, m_\tau,
m_Z, \alpha_3,
\alpha_{em}$ and $\sin^2\theta$ give boundary conditions at low energies which
constrain the
ten-dimensional parameter space at the unification scale.  Various strategies
for doing this
are considered in the next section.

\section{Solving the SSM}

In this section, I examine different strategies to solve the constraints on the
parameter
space of the SSM neglecting thresholds.  There are two basic strategies for
doing this: top-down and bottom-up.  Threshold effects greatly complicate
the problem and will be considered in the next section.

The top-down method chooses the ten parameters at the unification scale\break
$(M_X,\alpha_X,\lambda_b,\lambda_t, \lambda_\tau, m_{1/2}, m_0, A, \mu, B)$
and then computes all the
parameters in the Lagrangian at $m_Z$ using the renormalization group
equations.  Assuming
only real vev's of the neutral Higgs, the scalar potential at tree level is
$$V=(m^2_{\jbar H}+\mu^2)|\bar v|^2+(m^2_H+\mu^2)|v|^2+2B\mu\bar vv+{1\over
8}(g^2_2+{3\over 5}g^2_1)(|\bar v|^2-|v|^2)^2\eqno{(3.1)}$$
Extremizing this potential with respect to $\bar v$ and $v$ determines the
Higgs vev's,
which in turn determine $M_Z$ and $\tan\beta=\bar v/v$.
$$\eqalignno{
{2v\bar v\over v^2+\bar v^2}=\sin 2\beta &={-2B\mu\over (m^2_{\jbar
H}+m^2_H+2\mu^2)}\cr
{g^2_2\over{2\cos^2\theta}}(v^2+\bar v^2)=m^2_Z &={2(m^2_H-m^2_{\jbar H}
\tan^2\beta)\over
\tan^2\beta-1}-2\mu^2&{(3.2)}\cr}$$
An acceptable point in parameter space must also yield a potential bounded from
below with
the global minimum having non-zero vev's for only the two neutral Higgs.

For the points in the 10-d parameter space with acceptable minima, this process
determines
the entire SSM Lagrangian, including known experimental parameters like
$\alpha_3$,
$\sin^2\theta$, $\alpha_{em}$, $m_b$, $m_\tau$, $m_Z$.  These experimental
values constrain
the 10-d parameter space at the unification scale.
The top-down method may be computationally simplified by
expressing all the dimensionful parameters in units of
$m_{1/2}$.  Once the potential has been minimized to
determine $v/m_{1/2}$ and $\bar v/m_{1/2}$, the
constraint $m^2_W = g_2^2(v^2+\bar v^2)/2$ determines
$m_{1/2}$.  However when thresholds are included, the
dimensionful values of the parameters are required to
construct the potential, and this simplification is not
possible.

The bottom-up method attempts to simplify and speed up this process by directly
determining some of the parameters at the unification scale from known
low-energy parameters.
The gauge sector provides the classic example whereby, at one loop and
neglecting
thresholds, $\alpha_{em}$ and $\alpha_3$ predict $M_X$, $\alpha_X$, and
$\sin^2\theta$.
$$\eqalignno{
\sin^2\theta &={1\over{5}}+{{7\alpha_{em}}\over{15\alpha_3}}\cr
{1\over \alpha_X} &={3\over{20\alpha_{em}}}+{3\over{5\alpha_3}}\cr
\ln\Bigl({M_X\over m_Z}\Bigr)
&={\pi\over{30}}\biggl({3\over{\alpha_{em}}}-{8\over{\alpha_3}}\biggr)&{(3.3)}\cr}$$
The prediction of $\sin^2\theta$ represents a spectacular success of the SSM.
More careful
treatment of this prediction, including higher-loops and thresholds, has been
used to further
constrain SSM and SSM-like models.\bib{7}

The one-loop RGE's for the Yukawa couplings involve only the gauge couplings
and the
Yukawa couplings.  The fermion masses $m_b, m_t, m_\tau$ and $\tan\beta$ allow
$\lambda_b,
\lambda_t, \lambda_\tau$ to be determined at low energies and evolved to
calculate the
three Yukawas at the unification scale, independent of the soft
supersymmetric-breaking
parameters.  Since it is convenient to use $m_Z$ as the low-energy
renormalization point,
the physical fermion masses must be converted to running $\overline{MS}$ masses
renormalized at
$m_Z$.\bib{8}

Thus, from two unknowns, $m_t$ and $\tan\beta$, and the known fermion masses,
$m_b$ and
$m_\tau$, the Yukawas can be determined at $m_Z$ and evolved to the unification
scale.
Specifying the soft supersymmetry-breaking parameters $m_{1/2}, m_0,$ and $A$
at the
unification scale allows the entire set of RGE's (except the two for $\mu$ and
$B$ which
do not enter any of the other RGE's) to be evolved to determine all the
parameters at $m_Z$.
Having chosen $\tan\beta$ and knowing the experimental value of $m_Z$
determines the two
Higgs vev's through Eq.~(3.2).  The minimization conditions
$\partial V/\partial v = \partial V/\partial \bar v =0$
can now be solved for the remaining unknowns, $\mu$ and $B$, renormalized at
$m_Z$.  In this
method, the constraints from $m_b$, $m_\tau$, $m_Z$ (and
$m_t$ when it has been determined) are implemented at the
beginning of the procedure rather than at the end.  This
eliminates the computational problem encountered in the
top-down method of searching a large dimensional space
for a smaller dimensional subspace satisfying constraints.

The parameter space of the SSM is bounded in the $(m_t,\tan\beta)$ plane by a
combination of perturbative unitarity, the experimental lower bound on the top
mass,
and electroweak breaking.  The lower bound on the gluino mass bounds $m_{1/2}$
from
below, and naturalness bounds $m_{1/2}$ from above, as well as bounding the
magnitude of $\xi_0\equiv m_0/m_Z$ and $\xi_A\equiv A/m_Z$.  However the
naturalness bounds
require a precise definition of how much fine tuning is natural, and are
somewhat a matter of
taste.\bib{6,9}  Large five-dimensional areas of parameter space
are consistent with all known experimental constraints.

The lightest neutralino provides a natural dark matter candidate.  The SSM has
an exact R-parity
and the lightest supersymmetric particle (LSP) is stable.  Areas of parameter
space where the
lightest supersymmetric particle is a chargino or areas where the relic
neutralino density is
too great are excluded in the SSM,  although R-parity breaking extensions of
the SSM
could possibly rescue these areas.  However, cosmologically, the most
interesting areas of
parameter space are those where the LSP is a neutralino with relic density near
the critical
density.\bib{10}

\section{Thresholds}

The most straightforward method of including threshold
effects is to drop heavy particles from the RGE's at
scales below the mass of the heavy particles.  This
method has the advantage of maintaining the leading
log accuracy of the RGE's.  However, the method has two
computational drawbacks.  First, the RGE's and scalar
potential become substantially more complicated once a
non-supersymmetric threshold has been integrated out.\bib{11}
This is because the dimensionless parameters in the
scalar potential are no longer simply related to
gauge and Yukawa couplings by supersymmetry, leading
to a proliferation of parameters.  Second,
the threshold structure depends on the final solution, which
in general must be determined by
iteratively approaching a self-consistent solution.
Despite these two drawbacks, significant progress has
been made in determining and solving the
non-supersymmetric RGE's resulting from integrating out
the various thresholds in the SSM.\bib{12}

A computationally simpler approach uses the one-loop
corrections to the vacuum energy\bib{13} to construct a scalar
potential independent of the renormalization
point to one-loop order\bib{14}
$$\eqalignno{
V &=V_{tree} + \Delta V\cr
\Delta V &={1\over 64\pi^2} STr {\cal M}^4\biggl(ln{{\cal M}^2\over Q^2} - {3
\over 2}\biggr)&{(4.1)}\cr}$$
where $V_{tree}$ is given by Eq.~(3.1),
$STr~f({\cal M}^2)=\Sigma_j(-1)^{2j}(2j+1)Tr~f({\cal M}^2_j)$,
and ${\cal M}^2_j$ are the
field-dependent spin-j mass matrices.  The term $\Delta V$ incorporates
threshold effects to
the scalar potential and quantities derived from it (e.g. Higgs
masses).  The computation using the
bottom-up scheme is particularly simple.\bib{6}

\item{A) }Use $m_b(m_Z), m_t(m_Z), m_\tau(m_Z)$ and
$\tan\beta(m_Z)$ to compute $\lambda_1$, $\lambda_2$,
$\lambda_3$ at $m_Z$ and evolve up to $M_X$.

\item{B) }Choose $\xi_0\equiv m_0/m_{1/2}$, $\xi_A\equiv A/m_{1/2}$ at $M_X$
and evolve the RGE's down to $m_Z$.

\item{C) }For each choice of $m_{1/2}$, numerically solve the
equations $\partial V/\partial v = \partial V/\partial \bar v=0$ for $\mu$ and
$B$.
This method
requires only one iteration of the RGE's for each point in the ($m_t,
\tan\beta, \xi_0, \xi_A$)
parameter space and numeric solution of the minimization
equations for each value of $m_{1/2}$.  The procedure is
sufficiently efficient to search the whole
five-dimensional parameter space.  In addition, the
property that the scalar potential is independent of the
renormalization point to one-loop order provides a highly
non-trivial numerical check on the procedure.\bib{6}

The potential Eq.~(4.1) requires a slight adjustment to be
truly independent of the renormalization point:  as it
is, only derivatives of Eq.~(4.1) have this property.  Using
$\phi$ to denote the collection of field vev's and
$t=ln(Q/m_Z)$, the corrected potential is
$$V=V_{tree}(\phi, t) + \Delta V(\phi, t) - \Delta V(0,
t)\eqno{(4.2)}$$
This prescription can be justified in several ways.\bib{15}  Adding a
field-independent piece
to $V$ does not change the derivatives of $V$.  Furthermore, the
renormalized potential should resemble the tree level
potential with suitably modified coefficients and
wave-function renormalizations, which clearly vanishes at
the origin of field space as Eq.~(4.2) does.  Explicit
computation for a representative point in the SSM
parameter space confirms the one-loop renormalization
independence of Eq.~(4.2).
$${\partial V\over \partial t}\Big\vert_{t=0}=-0.0007m_{1/2}^4\eqno{(4.3)}$$
The departure of Eq.~(4.3) from zero is of the size expected
from round-off error in the single-precision computer
calculation used.  For comparison, a similar calculation
including only the stop and sbottom contributions in $\Delta V$
gives
$${\partial V\over \partial t}\Big\vert_{t=0}=-0.0224m_{1/2}^4\eqno{(4.4)}$$
The cancellation in Eq.~(4.3) is only observed when all
particles contributing to $\Delta V$ are included.

In numerically computing the derivative in Eqs.~(4.3) and (4.4),
the potential is computed at the renormalization points $m_Z$ ($t=0$) and
a renormalization point slightly higher than $m_Z$ ($t=\delta$).
Several subtle points necessary to numerically observe
the vanishing of $\partial V/\partial t$ are worth mentioning.  In
running the RGE's between $m_Z$ and $M_X$, and back, no
thresholds should be integrated out.  This would, in
effect, double count the influence of the threshold in
$V$:  once in $\Delta V$, and once by integrating out the
threshold in the RGE's.  The
masses in $\Delta V$ should be renormalized at $t=0$.  Note
that the potential is formally t-independent to one-loop
no matter where the masses in the supertrace are
renormalized.  However, only when these masses are
renormalized at $t=0$ do the two-loop subleading-log
terms and the linear t-term in the potential vanish.
In computing $V_{tree}$ at $t=\delta$, all the parameters in
$V_{tree}$ must be renormalized at $t=\delta$, including the wave-function
renormalization of the vev's $v$ and $\bar v$.
Two-loop and higher-loop effects (higher derivatives of
$V$ with respect to $t$) increase logarithmically with
$Q/m_Z$, and signal a progressive deterioration of the
one-loop approximation as the mass of the thresholds
increase.  Although it is difficult to formulate precise criteria for where the
approximation breaks down, higher-order terms certainly
become important for sparticle masses above about a $\TeV$.

Although the inclusion of threshold
effects shifts the detailed quantitative predictions of the SSM, the
qualitative
picture of the presently allowed regions in parameter space remains much the
same.
However, the tree-level result that $m_h<m_Z$ is substantially modified, and
the lightest
Higgs could be much heavier.\bib{16}
\vfill
\eject
\section{More Realistic Models}

Fundamental considerations clearly indicate that the
SSM is not a complete theory.
The gauge couplings in the SSM diverge above the unification scale.
Furthermore, quantum gravitational effects require
a fundamental change in the very framework of quantum field theory, which
the SSM is formulated in, at a scale not much larger than
the gauge unification scale.  GUT's provide a natural
structure to preserve the unification of gauge couplings
at scales larger than the gauge unification scale.
String theory provides a natural unified framework for
quantum field theory and gravity.

Supersymmetric $SU(5)$\bib{17} unifies the SSM gauge group within $SU(5)$.  The
fields added to
the SSM are superheavy X and Y gauge superfields, an extra adjoint chiral
superfield whose vev
breaks $SU(5)\to SU(3)\times SU(2)\times U(1)$, and $D^C, \jbar D^C$ chiral
superfields
filling out the $SU(5)$ representations containing the SSM Higgs
doublets.  Since all of these fields have masses near the gauge unification
scale, the
renormalizable Lagrangian at scales much below $M_X$ is that of the SSM with
couplings
slightly modified by GUT threshold corrections.  The fermionic components of
the superheavy
$D^C, \jbar D^C$ superfields mediate dimension-five, non-renormalizable,
proton-decay
operators.  Proton decay is absent in the SSM and would provide definite
evidence for
physics beyond the SSM.  Consideration of experimental limits on the proton
lifetime rule
out large areas of the supersymmetric $SU(5)$ parameter space.\bib{18}  The
results can be interpreted as showing that the model is either highly
predictive or highly
constrained, depending on one's point of view.
In addition, the group $SU(5)$ gives the generally successful prediction
$\lambda_\tau(M_X)=\lambda_b(M_X)$.\bib{19}  However, contour plots of $m_b$ in
the plane of
the most relevant parameters, $(\tan\beta, m_t)$,\bib{20} eliminate all but a
narrow strip
near the edge of the area of perturbative Yukawas.  The effect of less relevant
variables,
especially $\alpha_3$, must also be considered in a detailed analysis.\bib{21}

Essentially, testing the parameter space of supersymmetric $SU(5)$ reduces to
testing the
parameter space of the SSM with the addition of the proton decay constraint and
the constraint
$\lambda_b(M_X)=\lambda_{\tau}(M_X)$. Increasingly precise
measurements of $\sin^2\theta,\alpha_3, m_b$, and $\tau_P$ will either restrict
the
supersymmetric $SU(5)$ parameter space to unnaturally high supersymmetry
breaking masses or, if
$\tau_P$ is observed, impose two constraints on the five-dimensional parameter
space of the
SSM:  one from $\tau_P$, and one from $\lambda_b(M_X)=\lambda_\tau(M_X)$.
A theoretical problem with supersymmetric $SU(5)$ is the fine-tuning required
to split
light Higgs doublets from the superheavy Higgs triplets.  It is possible to
remedy
this problem in extensions of the model\bib{22} although these extensions
usually introduce
additional problems.  Another problem, in the context of the string, is that
the adjoint chiral
Higgs superfield needed to break $SU(5)$ cannot be obtained in the standard
$k=1$
construction.\bib{23}

Supersymmetric Flipped $SU(5)\times U(1)$\bib{24} gives an alternative GUT
comparable in
simplicity to supersymmetric $SU(5)$.  Since a ${\bf 10}$ and ${\bf
\overline{10}}$
representation breaks $SU(5)\times U(1)\to SU(3)\times SU(2)\times U(1)$,
Flipped $SU(5)\times
U(1)$ naturally arises in string constructions.\bib{25}  In addition, a missing
partner
mechanism splits the masses of the Higgs doublets and triplets without fine
tuning. Flipped
$SU(5)\times U(1)$ reduces to the SSM at scales much below $M_X$\bib{26} in
much the same way
as supersymmetric $SU(5)$.  In Flipped $SU(5)\times U(1)$, it can be shown that
the onset of
supersymmetry breaking, the gauge unification scale, and the $SU(5)\times
U(1)\to SU(3)\times
SU(2)\times U(1)$ scales must be within about an order of magnitude of each
other.\bib{26}
The departure from universal soft supersymmetry breaking in the resulting SSM
because of the
splitting between these three scales is therefore small.  Dimension-six
operators give the
dominant contribution to proton decay in Flipped $SU(5)\times U(1)$.
Experimental limits on
$\tau_P$ are just now reaching the sensitivity needed to begin to constrain the
model.\bib{27}
Because of the flipped embeddings, the relation
$\lambda_b(M_X)=\lambda_\tau(M_X)$ need
not hold in the flipped model, though this relation may be preserved in
string-derived
models due to an underlying $SO(10)$ symmetry.

The most promising ultimate explanation of the SSM is the string.  In addition
to
$SU(5)\times U(1)$ models, $SU(3)\times SU(2)\times U(1)$-like models naturally
emerge in
many string constructions.  Moreover, gauge coupling unification is a
prediction of string
models and the gauge unification scale is calculable in any particular
model.\bib{28}  A string
unification scale consistent with the SSM and low-energy measurements is
possible in
orbifold models with somewhat awkward choices of the moduli.\bib{29}  However,
for generic
values of the moduli, and for large classes of models constructed in the free
fermionic
formulation of the string, the string unification scale is too high, about
$10^{18}\GeV$.\bib{30}  A simple remedy to this problem is to add extra
vector-like
representations of chiral superfields with masses chosen to give the desired
string
unification scale while preserving the successful SSM prediction of
$\sin^2\theta$.  Using
only representations with Standard Model quantum numbers, the unique minimal
choice is an
extra $Q, \jbar Q$ pair with mass ${\cal O}(10^{13}\GeV)$ and an extra $D^C,
\jbar D^C$ pair
with mass ${\cal O}(10^5\GeV)$.  This minimal model is called the String
Inspired Standard
Model (SISM)\bib{31} and can be easily recast as a Flipped $SU(5)\times U(1)$
SISM\bib{32} by
adding an extra ${\bf 10, \overline{10}}$ pair of chiral superfields which
contain $Q,
\jbar Q$ and $D^C, \jbar D^C$ representations.  Many non-minimal choices exist
with larger
numbers of representations, though allowing only Standard Model quantum
numbers, at least
one $Q, \jbar Q$ pair is always needed.  This requirement can be relaxed by
allowing
representations with non-standard hypercharge.\bib{33}

Ultimately, string theory should predict the fermion masses and mixings.
Experimental
bounds have been shown to be consistent with extremely simple patterns for the
Yukawa
matrices at the unification scale.\bib{4}  These textures provide a reachable
target for
string models.  One approach attempts to fix the orbifold moduli to give the
observed
fermion masses and mixings.\bib{34}  String derivations of the soft
supersymmetry
parameters validate the assumption of predominantly universal soft
supersymmetry and add small
non-universal corrections.\bib{35}  In the top-down approach, each particular
string model
potentially fixes the high-energy boundary conditions.  In the bottom-up
approach, experiment
can be used to constrain the high-energy parameters and give an indication of
the classes of
string models required.  Considering the huge number of plausible string models
and the
computational effort required to derive predictions, an artful combination of
these
complementary approaches seems necessary.

\section{Sparticle Spectroscopy}

A remarkable consequence of the self-interacting dynamics
inherent in the framework of quantum field theories is
that every physical process depends on all the parameters
of the theory.  In principle, this means that for a
theory with n parameters, any n physical measurements
completely determine the theory, if these measurements
can be made with arbitrary precision.  However, in
practice, experimental measurements have a finite
precision, and different parameters in the theory enter
each process at different orders in the perturbative
expansion.  This makes the strategy of which processes
are used to determine which parameters important.

The most direct strategy is to actually produce the
particle of interest.  Definitive proof for supersymmetry
would be the production of sparticles.  If the masses of
several sparticles could be measured, the resulting
sparticle spectroscopy could be used to determine the
parameters of the SSM and search for departures from the
SSM.  The sparticles corresponding to the two light
generations have a very simple dependence on only three
of the SSM parameters: $m_0$, $m_{1/2}$, and $\tan\beta$.
$$m^2_{\tilde p} = m^2_0+c_{\tilde p}(m_{\tilde p})
m^2_{1/2} + 2\biggl[T^{\tilde p}_{3}-{3\over 5}Y^{\tilde
p}\tan^2\theta\biggr]m^2_W\cos 2\beta\eqno{(6.1)}$$
Measurements of three sparticle masses can be converted
into a determination of $m_{1/2}$ and $m_0$ with
fractional uncertainties comparable to that of the
sparticle mass measurements.\bib{36}

Sufficiently accurate
determination of more sparticle masses could be used to
discriminate between different extensions of the SSM such
as extensions of the gauge group, additional Yukawas,
generational-dependent extra heavy gauge bosons, and
non-universal sypersymmetry breaking, which all leave
distinct imprints on the sparticle spectrum.  For
example, for large enough sparticle masses
(so that the D-terms in Eq.~(6.1) can be neglected)
and $m_{\tilde g} = c_{\tilde g}m_{1/2}$
the quantity
$$\Delta_{ij}={m^2_i-m^2_j\over m^2_{\tilde g}} =
{c_i-c_j\over c^2_{\tilde g}}\eqno{(6.2)}$$
gives, e.g. $\Delta_{\tilde e_L\tilde e_R} = 0.062 (0.048)$
and $\Delta_{\tilde u_L\tilde d_R} = 0.088 (0.061)$ in the SISM
(SSM).\bib{31}  These measurements would discriminate between the
SISM and the SSM independent of the soft supersymmetry-breaking parameters
used.

\section{Flavor Changing Neutral Currents}

It is also possible to deduce information about particles
inaccessible directly at presently available energies
through their virtual influence in loop corrections to
precision experiments.  This is the strategy which has
been used to bound the top and Higgs masses by requiring
consistency of the different radiatively corrected
electroweak processes measured
at LEP.\bib{37}  Flavor changing neutral currents offer a particularly
sensitive probe of virtual loop effects.  Because flavor
changing neutral currents vanish at tree level, the
influence of the new superpartners can potentially be comparable to that of the
Standard
Model fields.

In calculating supersymmetric contributions to flavor changing neutral currents
by summing
over mass eigenstates, the final result is buried within cancellations between
the
contributions of six different squark mass eigenstates involving $6\times 6$
matrices of
quark-squark mixing angles.  An alternative, mass-insertion result\bib{38} may
be derived from
the mass-eigenstate result.  In the mass-eigenstate approach, the flavor
changing resides in
flavor non-diagonal gauge couplings while in the mass-insertion approach, the
gauge couplings
are flavor diagonal and the flavor changing resides in flavor off-diagonal
elements of the
sparticle mass matrices.  The difference between the two approaches is the
choice of flavor
basis and the result is independent of this choice.

For example, to lowest order in mass insertions, the gluino box contributions
to $\Delta
S=2$ processes gives rise to an effective interaction Lagrangian\bib{39}
$$\eqalignno{
  {\cal L}^{eff}_{\Delta S=2}&={\alpha^2_s\over 216M^2_{\tilde
q}}\Biggl\{\biggl({\delta\tilde
m_{\bar d_Ls_L}^2\over M_{\tilde q}^2}\biggr)^2
  [66\tilde f(x)+24xf(x)]
  (\bar d_i\gamma_{\mu}P_Ls_i)
  (\bar d_j\gamma^{\mu}P_Ls_j)\cr
&+\biggl( {\delta\tilde m_{\bar d_Rs_R}^2\over M_{\tilde q}^2}\biggr)^2
  [66\tilde f(x)+24xf(x)]
  (\bar d_i\gamma_{\mu}P_Rs_i)
  (\bar d_j\gamma^{\mu}P_Rs_j)\cr
&+\biggl({\delta\tilde m_{\bar d_Ls_L}^2\over M_{\tilde
q}^2}\biggr)\biggl({\delta\tilde m_{\bar
d_Rs_R}^2\over M_{\tilde q}^2}\biggr)
  \Bigl([-72\tilde f(x)+504xf(x)](\bar d_iP_Ls_i)(\bar d_jP_Rs_j)\cr
 &\hskip1.8in+[120\tilde f(x)+24xf(x)]
  (\bar d_iP_Ls_j)(\bar d_jP_Rs_i)\Bigr)\cr
&+\biggl({\delta\tilde m_{\bar d_Ls_R}^2\over M_{\tilde q}^2}\biggr)^2xf(x)
  \bigl[324(\bar d_iP_Rs_i)(\bar d_jP_Rs_j)-108(\bar d_iP_Rs_j)(\bar
d_jP_Rs_i)\bigr]\cr
&+\biggl({\delta\tilde m_{\bar d_Rs_L}^2\over M_{\tilde q}^2}\biggr)^2 xf(x)
  \bigl[324(\bar d_iP_Ls_i)(\bar d_jP_Ls_j)-108(\bar d_iP_Ls_j)(\bar
d_jP_Ls_i)\bigr]\cr
&+\biggl({\delta\tilde m_{\bar d_Ls_R}^2\over M_{\tilde
q}^2}\biggr)\biggl({\delta\tilde m_{\bar
d_Rs_L}^2\over M_{\tilde q}^2}\biggr) \tilde f(x)
  \bigl[108(\bar d_iP_Ls_i)(\bar d_jP_Rs_j)\cr
 &\hskip2.1in -324(\bar d_iP_Ls_j)(\bar d_jP_Rs_i)\bigr]\Biggr\}&{(7.1)}\cr}$$
with
$$\eqalignno{
f(x)&={1\over6(1-x)^5}(-6\ln{x}-18x\ln{x}-x^3+9x^2+9x-17)&{(7.2a)}\cr
\tilde
f(x)&={1\over3(1-x)^5}(-6x^2\ln{x}-6x\ln{x}+1x^3+9x^2-9x-1)&{(7.2b)}\cr}$$
where $M_{\tilde q}$ is the universal (or average) down-squark mass and
$x=M^2_{\tilde g}/M^2_{\tilde q}$.  The $\Delta B=2$ effective interaction can
be
read directly from Eq.~(7.1) by the substitution $s\to b$.  Requiring that each
chiral
contribution not exceed the experimental value of $\Delta M_K$, $\Delta M_B$
and
$\epsilon_K$ gives upper bounds on the various mass insertions summarized in
Table 1:
\midinsert{\tabskip=0pt \offinterlineskip
\def\tablerule{\noalign{\hrule}}%
\def\struta{\vrule height13pt depth 6pt width 0pt}%
\def\strutb{\vrule height16pt depth 11pt width 0pt}%
\def\strutc{\vrule height20pt depth 11pt width 0pt}%
\vskip-\abovedisplayskip
$$\vcenter{\halign to 418pt{\struta#& \vrule#\tabskip=1em plus2em&
  \hfil#\hfil& \vrule#&\hfil#\hfil& \vrule#&\hfil#\hfil&
  \vrule#&\hfil#\hfil& \vrule#\tabskip=0pt\cr\tablerule
\strutb&&\multispan7\hfil Phenomenological Upper Bounds on \ds and \db
\hfil&\cr\tablerule
\strutc&&\omit&&\multispan3\hfil
   $\Bigl({\delta\tilde m^2_{\jbar d_As_B}\over M^2_{\tilde q}}\Bigr)
   \Bigl({\delta\tilde m^2_{\jbar d_{A'}s_{B'}}\over M^2_{\tilde q}}\Bigr)$
   \hfil&&
   $\Bigl({\delta\tilde m^2_{\jbar d_Ab_B}\over M^2_{\tilde q}}\Bigr)
   \Bigl({\delta\tilde m^2_{\jbar d_{A'}b_{B'}}\over M^2_{\tilde q}}\Bigr)$
   \hfil&\cr\tablerule
&&$(\alpha\beta)(\alpha'\beta')$&&$\Delta M_K$&&$\epsilon_K$&&$\Delta M_B$&\cr
\tablerule
&&$(LL)^2$ or $(RR)^2$&&$(0.10)^2$&&$(0.0080)^2$&&$(0.27)^2$&\cr\tablerule
&&$(LL)(RR)$&&$(0.0060)^2$&&$(0.00049)^2$&&$(0.073)^2$&\cr\tablerule
&&$(LR)^2$ or $(RL)^2$&&$(0.0082)^2$&&$(0.00066)^2$&&$(0.082)^2$&\cr\tablerule
&&$(LR)(RL)$&&$(0.044)^2$&&$(0.0035)^2$&&$(0.14)^2$&\cr\tablerule}}$$
\vskip-\belowdisplayskip
\vskip.05in{\leftskip=.2truein\rightskip=\leftskip\baselineskip=12pt\tenpoint\noindent TABLE I.
All numbers in the table must be multiplied by the factor $(M_{\tilde q}/{\rm
TeV})^2$.  Numerical
values assume $\sqrt{x}\sim M_{\tilde g}/M_{\tilde q}=1$.  Stricter (weaker)
results generally
apply to $x<1$ ($x>1$).  Bounds derived from $\epsilon_K$ assume maximal CP
violation.  Bounds
from $\Delta M_B$ must be scaled by (160 MeV/$f_B$).\par}}
\endinsert
\noindent
All the constraints are useful (i.e. $\delta\tilde m^2_{AB}/M^2_{\tilde q}<1$),
even for
heavy squark masses $M_{\tilde q}\gg 1\TeV$.  This severely restricts the
flavor structure
of a general supersymmetric theory.

In the SSM, non-diagonal squark mass matrices are generated by renormalization
effects between
the unification scale and $M_Z$.  The resulting low-energy mass insertions are
predominantly
left-left, and exhibit the following approximate flavor dependence:\bib{39}
$${[\delta\tilde m^2_{\bar d_Ld_L}]_{ij}\over M_{\tilde q}^2}
  =c_{LL}(\xi_0,\xi_A)[K^\dagger\lambda_u^2 K]_{ij}\eqno{(7.3)}$$
where $\lambda_u$ is the diagonalized charge $2/3$ quark Yukawa matrix and $K$
is the KM
matrix.  The function $c_{LL}(\xi_0,\xi_A)$ may be calculated by running the
one-loop
matrix RGE's and one finds that $\vert c_{LL}(\xi_0,\xi_A)\vert < 1$ for
$\vert\xi_0\vert$,
$\vert\xi_A\vert < 5$.  Combining this with bounds on the various KM matrix
elements
reveals that the SSM safely satisfies all the bounds in Table 1.

The recent CLEO observation of $B\to K^*\gamma$\bib{40} suggests that $b\to
s\gamma$ may
severely constrain the SSM.  In the Standard Model with two Higgs doublets,
calculations of
$b\to s\gamma$ bound the mass of the charged Higgs.\bib{41}  However, in the
limit of exact supersymmetry, the chargino contributions cancel the two Higgs
SM contributions
and $BR(b\to s\gamma)=0$.\bib{42}  Despite the complexities of a full
calculation in the SSM
including two-loop QCD corrections, the resulting operator mixing, and all the
supersymmetric
contributions, preliminary calculations reveal that $b\to s\gamma$ is a very
promising
constraint on the parameters of the SSM.\bib{43}

Most nontrivial extensions of the SSM contain extra Yukawa couplings, and these
generically lead to large FCNC's.\bib{44}  In supersymmetric Flipped
$SU(5)\times
U(1)$,\bib{24} for example, there can be potentially large FCNC's generated
above the GUT
scale:\bib{39,45}
$$\eqalignno{ [\delta\tilde m_{\bar d_L d_L}^2]_{ij}
&=-{1\over 8\pi^2}[\lambda_6^*\lambda_6^T]_{ij}
\ln\biggl({M_{Pl}\over M_{GUT}}\biggr)[3m^2_0+A^2]&{(7.4a)}\cr
\noalign{\hbox{}}
[\delta\tilde m_{\bar d_R d_R}^2]_{ij}
&=[\delta\tilde m_{\bar d_L d_L}^2]_{ij}^*&{(7.4b)}\cr
\noalign{\hbox{}}
[\delta\tilde m_{\bar d_R d_L}^2]_{ij}
&=-\eta^{1}_{ij}v={vA\over 8\pi^2}
[\lambda_6\lambda_6^\dagger\lambda_d
+\lambda_d\lambda_6^*\lambda_6^T]_{ij}\ln\biggl({M_{Pl}\over
M_{GUT}}\biggr)&{(7.4c)}\cr
\noalign{\hbox{}}
[\delta\tilde m_{\bar d_L d_R}^2]_{ij}
&=[\delta\tilde m_{\bar d_R d_L}^2]_{ij}^*
&{(7.4d)}\cr}$$
where $\lambda_6$ is an {\it a-priori} unknown Yukawa coupling associated with
a see-saw neutrino
mass mechanism.

 From the phenomenological constraints in Table I, useful bounds on
the unknown GUT Yukawa $\lambda_6$ result.  The strongest bounds come from the
$(\delta\tilde
m^2_{LL})(\delta\tilde m^2_{RR})$ contributions to $\Delta M_K$, $\Delta M_B$:
$$\eqalignno{
\bigl|[\lambda_6^*\lambda_6^T]_{12}\bigr|\ln{\Bigl({M_{Pl}\over M_{GUT}}\Bigr)}
&<0.47{\xi_0^2+6\over 3\xi_0^2+\xi_A^2}\Bigl({M_{\tilde q}\over 1\TeV}\Bigr)
&{(7.5a)}\cr
\noalign{\hbox{}}
\bigl|[\lambda_6^*\lambda_6^T]_{13}\bigr|\ln{\Bigl({M_{Pl}\over M_{GUT}}\Bigr)}
&<1.28{\xi_0^2+6\over 3\xi_0^2+\xi_A^2}\Bigl({M_{\tilde q}\over 1\TeV}\Bigr)
&{(7.5b)}\cr}$$
using the approximate low-energy relation $M^2_{\tilde q}\simeq (\xi^2_0+6)
m^2_{1/2}$.
Equation (7.5) provides important knowledge
about the unknown matrix $\lambda_6$ and the pattern of soft supersymmetry
breaking.
For instance, if supersymmetry breaking takes the form of either $m_0$ or A,
then Eq.~(7.5)
requires certain elements of $\lambda_6$ to be quite small---smaller than
expected from
superstring theories, which relate Yukawa couplings to gauge couplings
$\lambda\approx
g\approx 0.7$. One may be forced to conclude that these couplings vanish at the
tree level
in such theories---or that soft
supersymmetry breaking takes the form of $m_{1/2}$.
Note that because of Eq.~(7.4b,d), the Flipped contribution to CP violation
$\epsilon_K$
vanishes.  Flipped $SU(5)\times U(1)$ can, however, contribute significantly to
``direct" CP
violation $\epsilon'$, $K_L\to\pi^0ee$ and $K_L\to\pi^0\nu\bar\nu$, as shown in
ref. 39.

Perhaps the most interesting process in Flipped $SU(5)\times U(1)$ is $\mu\to
e\gamma$. A
non-trivial constraint on the elements of the arbitrary unitary matrix $U$,
which embeds the
lepton doublets along with the up-quark triplets in the GUT representations,
may be derived
from the limits on $BR(\mu\to e\gamma)$:\bib{39}
$$|U^*_{13}U_{23}|\ln{\Bigl({M_{Pl}\over
M_{GUT}}\Bigr)}\lsim{3.6\times10^{-3}\over
\lambda^2_t} \Bigl({0.276\over E(x_{\tilde B})}\Bigr)
\Bigl({M_{\tilde l}\over 100\GeV}\Bigr)^2
\sqrt{{BR(\mu\to e\gamma)\over 4.9\times10^{-11}}}\eqno{(7.6)}$$
where $x_{\tilde B}=m^2_{\tilde B}/M^2_{\tilde l}$
and $E(1)=.276$. The constraints from $BR(\tau\to \mu\gamma)$ and $BR(\tau\to
e\gamma)$ are
not interesting at present experimental limits.
A distinctive signature of Flipped $SU(5)\times U(1)$ is that the final-state
electrons and
photons are always left-handed. In conventional supersymmetric $SU(5)$ the
final-state
electrons and photons are always right-handed, but since the unknown matrix $U$
appearing in
Eq.~(7.6) gets replaced by the $KM$ matrix, the branching ratio for $\mu\to
e\gamma$ is
suppressed well below presently accessible levels.

\section{Conclusions}

The SSM constitutes a simple, well-defined basis for testing unified
supersymmetric
models.  The experimental predictions of the SSM are unfolded over a
five-dimensional parameter space $(m_t,\tan\beta,m_{1/2},m_0,A)$.  Large
regions of
this parameter space are consistent with all known experimental results.  Two
complementary methods of solving the SSM are the top-down and bottom-up
methods.
Thresholds effects may be included in evolving the RGE's or in one-loop
corrections
to the vacuum energy.  More realistic models like supersymmetric $SU(5)$,
Flipped
$SU(5)\times U(1)$, the SISM, and string-derived models are well approximated
by
the SSM.  In addition, more realistic models can give rise to processes like
proton
decay or lepton flavor violation which are forbidden in the SSM, as well as
constraints on the parameter space of the SSM like
$\lambda_b(M_X)=\lambda_{\tau}(M_X)$.
String-derived models are beginning to give explicit predictions for the Yukawa
couplings and soft supersymmetry-breaking parameters.  Sparticle spectroscopy
and flavor
changing neutral currents are examples of tests which might determine the
parameters of the SSM and provide a means to discriminate between more
realistic
models.

Despite the simplicity of the SSM, it is still an active area of research.
Only in
the last few years has its constrained parameter space been identified and
searched.  Combining the merits of the two methods of treating thresholds,
complete
two-loop calculations, and estimation of the accuracy of one-loop and two-loop
treatments all seem possible and worthwhile projects.

The interplay between the SSM, string-inspired, and string-derived models is
very lively.
Perhaps the most exciting prospect arises from the glimpse that string theory
gives into
physics beyond the Planck scale, physics beyond space-time.  The numerous
possible string
vacua suggest a symmetric state of string theory (perhaps topological field
theory) where
the space-time metric vanishes or is singular. Thus it seems that the ultimate
explanation
and source of the laws of nature transcends space and time.  But the scientific
method has
developed and operates within space and time.  However, purely empirical
science rests on a
foundation of theoretical insight  where theories are tested on the basis of
elegance and
logical consistency. The need for a quantum consistent theory of gravity
provides the most
compelling motivation for string theory.  In the search for ultimate truth,
especially in
the high-energy realms where experimental evidence is scarce, we may need to
sharpen our
skills for directly cognizing the structure of natural law.

\section{References}
\raggedright
\leftskip=1.2\parindent\parindent=0pt
  \vskip0pt plus.07\vsize\penalty-250\vskip0pt plus-.07\vsize

\rfrnc
J.~Ellis, S.~Kelley and D. V.~Nanopoulos, \PRL\ {\bf 249} (1990) 441;\break
P.~Langacker and M.~Luo, {\it Phys.~Rev.} {\bf D44} (1991) 817;\break
U.~Amaldi, W.~de Boer and H.~F\"urstenau, \PL\ {\bf 260B} (1991) 447.//

\rfrnc
L. J. Hall and U.~Sarid, LBL preprint LBL-32905 (1992).//

\rfrnc
J. P. Derendinger and C.A.~Savoy, \NP\ {\bf B237} (1984) 307;\break
J.~Ellis, J. S.~Hagelin, S.~Kelley and D. V.~Nanopoulos, \NP\ {\bf
B311}\nextline 1988 1.//

\rfrnc
H.~Georgi and C.~Jarlskog, \PL\ {\bf 86B} (1979) 297;\break
S.~Dimopoulos, L. J.~Hall and S.~Raby, \PRL\ {\bf 68} (1992) 1984.//

\rfrnc
S.~Bertolini, F.~Borzumati, A.~Masiero, and G.~Ridolfi, \NP\ {\bf B353} (1991)
591.//

\rfrnc
S.~Kelley, J. L.~Lopez, D. V.~Nanopoulos, H.~Pois and K.~Yuan, \NP\ {\bf B398}
(1993) 3.//

\rfrnc
J.~Ellis, S.~Kelley and D. V.~Nanopoulos, \PL\ {\bf 260B} (1991) 131;\break
F.~Anselmo, L.~Cifarelli, A.~Peterman and A.~Zichichi, {\it Nuovo Cimento} {\bf
105A} (1992)
581;\break
P.~Langacker and N.~Polonsky, \PR\ {\bf D47} (1993) 4028.//

\rfrnc
S.~Kelley, J. L.~Lopez and D. V.~Nanopoulos, \PL\ {\bf 274B} (1992) 387.//

\rfrnc
R.~Barbieri and G.~Giudice, \NP\ {\bf B306} (1988) 63.//

\eject
\rfrnc
J.~Ellis, D. V.~Nanopoulos, K. A.~Olive, and M.~Srednicki, \NP\ {\bf B238}
(1984)
453;\nextline S.~Kelley, J. L.~Lopez, D. V.~Nanoupolos, K.~Yuan, and H.~Pois,
{\it
Phys.~Rev.} {\bf D47} (1993) 2461.//

\rfrnc
P. H.~Chankowski, {\it Phys.~Rev.} {\bf D41} (1990) 2877.//

\rfrnc
H. E.~Haber and R.~Hempfling, SCIPP 91/33 (1992).//

\rfrnc
S.~Coleman and E.~Weinberg, {\it Phys.~Rev.} {\bf D7} (1973) 1888.//

\rfrnc
G.~Gamberini, G.~Ridolfi, and F.~Zwirner, \NP\ {\bf B331} (1990) 331.//

\rfrnc
B.~Kastening, \PL\ {\bf 283B} (1992) 287.//

\rfrnc
J.~Ellis, G.~Ridolfi, and F.~Zwirner, \PL\ {\bf 257B} (1991) 83.//

\rfrnc
S.~Dimopoulos and H.~Georgi, \NP\ {\bf B193} (1981) 150; \nextline
N.~Sakai, {\it Z.~Phys.} {\bf{C11}} (1982) 153;\nextline
A.~Chamseddine, R.~Arnowitt and P.~Nath, \PRL\ {\bf 105} (1982) 970.//

\rfrnc
B.~Campbell, J.~Ellis and D. V.~Nanopoulos, \PL\ {\bf 141B} (1984) 229;
\nextline
R.~Arnowitt and P.~Nath, \PR\ {\bf D38} (1988) 1479.//

\rfrnc
M. S.~Chanowitz, J.~Ellis, and M. K.~Gaillard, \NP\ {\bf B128} (1977)
506;\nextline
A. J.~Buras, J.~Ellis, M. K.~Gaillard, and D. V.~Nanopoulos, \NP\ {\bf B315}
(1978)
66;\nextline D. V.~Nanopoulos, and D. A.~Ross, \NP\ {\bf B157} (1979) 273; \PL\
\nextline{\bf
108B} (1982) 351; \PL\ {\bf 118B} (1982) 99.//

\rfrnc
J.~Ellis, S.~Kelley, and D. V.~Nanopoulos, \NP\ {\bf B373} (1992) 55.//

\rfrnc
P.~Langacker and N.~Polonsky, University of Pennsylvania preprint UPR-0556T
(1993).//

\rfrnc
A.~Masiero, D. V.~Nanopoulos, K.~Tamvakis, and T.~Yanagida, \PL\ {\bf 115B}
(1982) 380;\nextline
B.~Grinstein, \NP\ {\bf B206} (1982) 387;\nextline
K.~Inoue, A.~Kakuto and T.~Tankano, {\it Prog. Theor. Phys.} {\bf 75} (1986)
664;\nextline
R.~Barbieri, G.~Dvali and A.~Strumia, \NP\ {\bf B391} (1993) 487.//

\rfrnc
H.~Dreiner, J. L.~Lopez, D. V.~Nanopoulos, and D.~Reiss, \PL\ {\bf 216B} (1989)
289.//

\rfrnc
I.~Antoniadis, J.~Ellis, J. S.~Hagelin, and D. V.~Nanopoulos, \PL\ {\bf 194B}
(1987) 231.//

\rfrnc
I.~Antoniadis, J.~Ellis, J. S.~Hagelin, and D. V.~Nanopoulos, \PL\ {\bf 231B}
(1989) 65.//

\rfrnc
J.~Ellis, J. S.~Hagelin, S.~Kelley, and D. V.~Nanopoulos, \NP\ {\bf B311}
(1988) 1.//

\rfrnc
J. S.~Hagelin and S.~Kelley, MIU preprint MIU-THP-61/92 (1993).//

\rfrnc
J.~Minahan, \NP\ {\bf B298} (1988) 36;\nextline
V. S.~Kaplunovsky, \NP\ {\bf B307} (1988) 145.//

\rfrnc
L. E.~Ibanez, D.~Lust, and G. G.~Ross, \PL\ {\bf 272B} (1991) 251.//

\rfrnc
I.~Antoniadis, J.~Ellis, S.~Kelley, and D. V.~Nanopoulos, \PL\ {\bf 272B}
(1991) 31.//

\rfrnc
S.~Kelley, J. L.~Lopez, D. V.~Nanopoulos, \PL\ {\bf 278B} (1992) 140.//

\rfrnc
J. L.~Lopez, D. V.~Nanopoulos, and A.~Zichichi, CERN-TH.6903/93 (1993).//

\rfrnc
A. E.~Faraggi, \PL\ {\bf 302B} (1993) 202.//

\rfrnc
B.~Carlos, F.~Gomez, and C.~Mu\~noz, \PL\ {\bf 292B} (1992) 42.//

\rfrnc
B.~Carlos, J. A.~Casas, and C.~Mu\~noz, CERN-TH.6681/92 (1992);\nextline
V. S.~Kaplunovsky and J.~Louis, \PL\ {\bf 306B} (1993) 269.//

\rfrnc
A. E.~Faraggi, J. S.~Hagelin, S.~Kelley, and D. V.~Nanopoulos, \PR\ {\bf D45}
(1991) 3272.//

\rfrnc
J.~Ellis and F. L.~Fogli, \PL\ {\bf 232B} (1989) 139;\nextline
P.~Langacker, \PRL\ {\bf 63} (1989) 1920.//

\rfrnc
M.~Dugan, B.~Grinstein, and L. J.~Hall, \NP\ {\bf B255} (1985) 413.//

\rfrnc
J. S.~Hagelin, S.~Kelley, and T.~Tanaka, MIU preprint MIU-THP-59/92.//

\rfrnc
E.~Thorndike, CLEO Collab., talk given at the 1993 Meeting of the American
Physical Society,
Washington D.C., April 1993.//

\rfrnc
V.~Barger, M.~Berger, and R. J.~Phillips, \PRL\ {\bf 70} (1993) 1368;\nextline
J.~Hewett, \PRL\ {\bf 70} (1993) 1045.//

\rfrnc
R.~Barbieri and G.~Giudice, CERN-TH.6830/93.//

\rfrnc
J. L.~Lopez, D. V.~Nanopoulos, and G. T.~Park, Texas A \& M University preprint
CTP-TAMU-16/93 (1993).//

\rfrnc
L. J.~Hall, V .A.~Kostelecky, and S.~Raby, \NP\ {\bf B267} (1986) 415.//

\rfrnc
F.~Gabbiani and A.~Masiero, \PL\ {\bf 209B} (1988) 289;\nextline
A.~Faraggi, J. L.~Lopez, D. V.~Nanopoulos, and K.~Yuan, \PL\ {\bf 221B} (1989)
337.//\par

\end